\documentclass[%
 reprint,
    superscriptaddress,
    amsmath,amssymb,
    aps,prl
]{revtex4-2}
\AtBeginDocument{\RenewCommandCopy\qty\SI}

\usepackage{graphicx}
\usepackage{dcolumn}
\usepackage{bm}
\usepackage{physics}
\usepackage{color}
\usepackage{pstricks} 
\usepackage{cancel} 
\usepackage{ulem} 
\usepackage[version=4]{mhchem}
\usepackage{siunitx}

\begin{document}

\title{Strong‑Field Photoelectron Interferometry with Near‑Single‑Cycle Yb Lasers}

\author{Mahmudul Hasan}
\email{These authors contributed equally to this work}
\affiliation{James R. Macdonald Laboratory, Department of Physics, Kansas State University, Manhattan, Kansas, 66506, USA}

\author{Phi-Hung Tran}
\email{These authors contributed equally to this work}
\affiliation{Department of Physics, University of Connecticut, Storrs, Connecticut 06268, USA}

\author{Jingsong Gao}
\affiliation{James R. Macdonald Laboratory, Department of Physics, Kansas State University, Manhattan, Kansas, 66506, USA}

\author{Van-Hung Hoang}
\affiliation{James R. Macdonald Laboratory, Department of Physics, Kansas State University, Manhattan, Kansas, 66506, USA}

\author{Ming-Shian Tsai}
\affiliation{Institute of Photonics Technologies, National Tsing Hua University, Hsinchu 300044, Taiwan}

\author{Ming-Chang Chen}
\affiliation{Institute of Photonics Technologies, National Tsing Hua University, Hsinchu 300044, Taiwan}

\author{Uwe Thumm}
\affiliation{James R. Macdonald Laboratory, Department of Physics, Kansas State University, Manhattan, Kansas, 66506, USA}

\author{Charles Lewis Cocke}
\affiliation{James R. Macdonald Laboratory, Department of Physics, Kansas State University, Manhattan, Kansas, 66506, USA}

\author{Chii-Dong Lin}
\affiliation{James R. Macdonald Laboratory, Department of Physics, Kansas State University, Manhattan, Kansas, 66506, USA}

\author{Anh-Thu Le}
\email{thu.le@uconn.edu}
\affiliation{Department of Physics, University of Connecticut, Storrs, Connecticut 06268, USA}

\author{Meng Han} 
\email{meng9@ksu.edu}
\affiliation{James R. Macdonald Laboratory, Department of Physics, Kansas State University, Manhattan, Kansas, 66506, USA}

\date{\today}

\begin{abstract}
The concept of using photoelectron interferometry in short laser fields to probe electron dynamics and target structures was introduced more than two decades ago. However, the quality of experimental data has remained insufficient for quantitative analysis, largely due to the instability of few-cycle Ti:Sa laser pulses—the current workhorse of short pulses. Here, we report the first systematic strong-field ionization experiments performed with industrial-grade, carrier-envelope-phase (CEP) stabilized, near-single-cycle Yb lasers. By measuring photoelectron momentum distributions in the direct-ionization regime, we show that single-cycle cosine-shaped pulses can separate and enhance both spider-leg and fishbone holographic structures. The spider-leg structure enables extraction of the electron scattering phase from the Ar atomic potential—information typically accessible only through attosecond metrology—while the fishbone structure reveals the orbital-parity contrast between Ar atoms and nitrogen molecules. Our measurements are quantitatively reproduced by both semiclassical Herman-Kluk-propagator and \textit{ab initio} simulations, paving the way for precision studies of electron–molecule scattering with widely accessible industrial-grade lasers.
\end{abstract}

\maketitle

Strong-field tunneling ionization plays a central role in both strong-field physics and attosecond science \cite{corkum2007attosecond}. The photoelectron momentum distribution (PMD) generated through tunneling ionization encodes rich information about laser-driven electron dynamics \cite{han2018attoclock,han2017revealing,eckart2018ultrafast,hartung2019magnetic,han2021complete,orunesajo2025phase} as well as the structure of the parent atoms or molecules \cite{morishita2008accurate,blaga2012imaging,wolter2016ultrafast}. In the direct-ionization regime, however, the coexistence of multiple interference and diffraction patterns makes this information difficult to extract. For instance, inter-cycle \cite{arbo2010intracycle,arbo2010diffraction} and intra-cycle \cite{gopal2009three,xie2012attosecond,richter2015streaking} interferences give rise to above-threshold ionization (ATI) rings and temporal double-slit fringes, respectively. Moreover, interference between direct and rescattered electrons generates holographic features \cite{de2020all,bian2011subcycle}, including spider-leg-like \cite{huismans2011time} and fishbone-like \cite{haertelt2016probing} patterns. Over the past two decades, it has become widely recognized that employing extremely short laser pulses—approaching the single-cycle limit—can limit rescattering to a single event, thereby greatly simplifying the PMD. Several pioneering experiments \cite{kling2008imaging,gopal2009three,khurelbaatar2024strong} have pursued this strategy. Nonetheless, none have succeeded in quantitatively extracting the electron scattering phase to compare to theory. This phase is a fundamental quantity characterizing the photoemission dynamics, and its energy derivative is directly related to the Wigner time delay \cite{zhang2011streaking,dahlstrom2013theory,pazourek15a,thumm2015fundamentals,shi2025photoelectron}. The inability of strong-field ionization to provide such quantitative access is largely due to poor data quality, stemming from the notorious instability of few-cycle Ti:Sa laser pulses.

In recent years, industrial-grade Yb-based lasers have attracted considerable attention for their robustness, high repetition rates, and high average power. However, their native pulse durations—typically longer than 100 fs—render them unsuitable for strong-field ionization. This limitation can be overcome through post-compression techniques \cite{jeong2018direct,lu2019greater,lavenu2018nonlinear,beetar2020multioctave,tsai2022nonlinear}, which shorten the pulses into the few-cycle or even single-cycle regime. In this Letter, we present the first strong-field ionization experiments performed with CEP-stabilized, sub-4-fs Yb lasers operating at a 10 kHz repetition rate. Using a velocity map imaging (VMI) spectrometer, we measured the PMDs of Ar atoms and nitrogen molecules—two targets with comparable ionization potentials—at different CEPs. We observe that direct-ionization electrons exhibit a pronounced streaking effect in momentum space, with the total ionization yield maximized for cosine-shaped pulses. Crucially, the high quality of our data allows us to extract the electron scattering phase from spider-leg structures and to identify the orbital-parity effect in fishbone structures, both in quantitative agreement with ab initio simulations. This work further benchmarks a recently developed semiclassical strong-field Herman-Kluk (SFHK) method \cite{tran2024quantum}. Unless otherwise specified, “sine-shaped” and “cosine-shaped” refer to the electric field waveform.


\begin{figure}[htbp!]
\centering
\includegraphics[width=\linewidth]{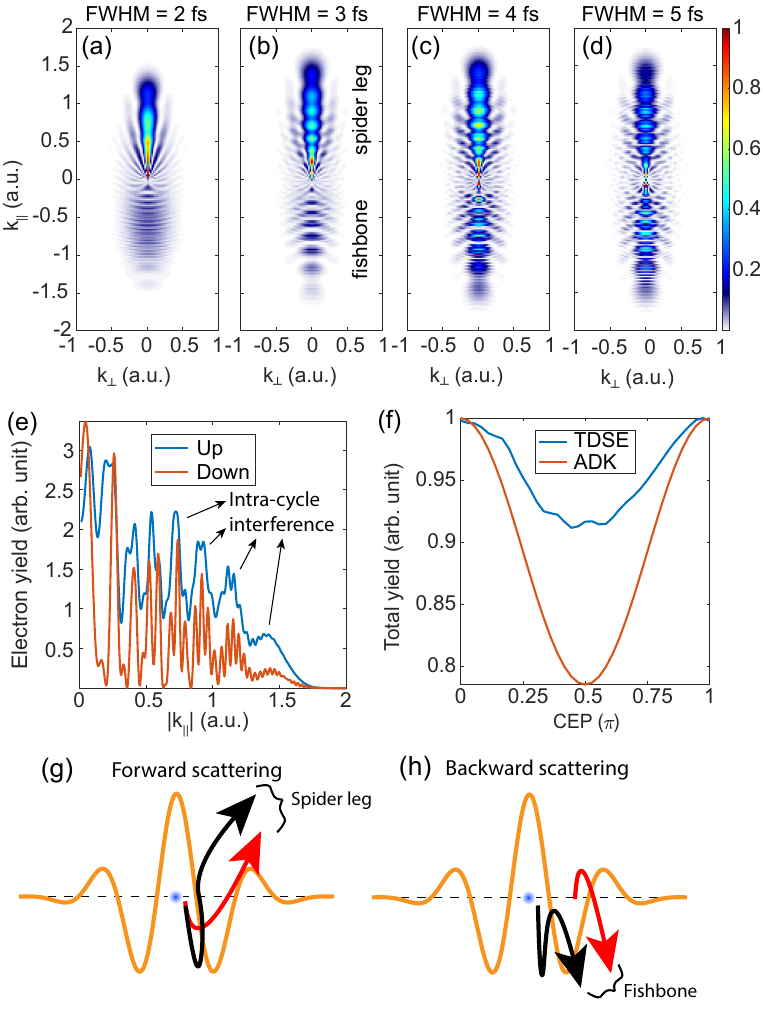}
\caption{(\textbf{a-d}) Calculated PMDs by solving the TDSE for a cosine-shaped few-cycle pulse with durations of 2, 3, 4, and 5 fs, respectively. The results indicate that the up-down asymmetry, i.e., the streaking effect on direct ionization electrons, becomes significantly weaker when the pulse duration exceeds 5 fs. (\textbf{e}) Lineouts from (b), showing photoelectron momentum distributions along the light polarization direction (vertical axis). (\textbf{f}) Calculated total ionization yield for a 3-fs pulse. The CEP of zero corresponds to a cosine-shaped pulse, which maximizes the ionization yield. The total ionization yield exhibits a $\pi$-periodicity with respect to CEP, meaning that cosine-shaped and -cosine-shaped pulses produce the same total ionization. Note that the CEP dependence of the total ionization yield rapidly diminishes as the pulse duration increases. (\textbf{g, h}) Schematics of two types of photoelectron holography, which can be separated in momentum space and are enhanced with cosine-shaped single-cycle pulses.} 
\label{fig:figure1}
\end{figure}

It is well established that the CEP of a short laser pulse influences the asymmetry of PMDs in the rescattering-dominated regime ($E > 2U_p$) \cite{paulus2001absolute,paulus2003measurement,wittmann2009single}, where $E$ is the photoelectron kinetic energy and $U_p$ is the ponderomotive energy of the laser field. In contrast, photoelectron interferometric structures typically arise in the direct-electron regime ($E < 2U_p$), where CEP effects are less explored and more difficult to observe. In Fig. 1, we first establish the requirements for observing CEP dependence in this regime. Using time-dependent Schrödinger equation (TDSE) simulations with a single-active-electron potential \cite{tong2005empirical}, we calculate PMDs from Ar atoms for pulse durations of 2, 3, 4, and 5 fs, shown in Figs. 1(a–d). The pulse duration is defined as the full width at half maximum (FWHM) of a Gaussian-shaped intensity envelope. The laser intensity is $2 \times 10^{14} \ \rm{W/cm}^{2}$, with a center wavelength of 900 nm (corresponding to an optical period of 3 fs), matching our experimental conditions. The light field is linearly polarized along the vertical axis. The calculated distributions in Figs. 1(a–d) primarily represent direct-ionization electrons. A pronounced up–down asymmetry appears for 2–4 fs pulses, but diminishes markedly at 5 fs (i.e., 1.7 optical cycles). Figure 1(e) shows the momentum distribution along the polarization axis for a 3-fs pulse (1 optical cycle): the widely spaced fringes correspond to intra-cycle interference, while the denser fringes reflect inter-cycle interference.

Figure 1(f) presents the CEP-resolved total ionization yield after integrating over full emission angle and full energy range. In addition to the TDSE results, we perform a simple calculation based on the ADK rate \cite{delone1991energy}, accounting only for direct electrons. Both calculations show that a cosine-shaped pulse (CEP = 0) results in greater ionization than a sine-shaped pulse. The total yield exhibits a $\pi$-periodicity in its CEP dependence, in contrast to the $2\pi$-periodicity in the spatial asymmetry of the high-energy rescattering electrons \cite{paulus2003measurement,wittmann2009single}. When the pulse duration is longer than 4 fs, the CEP dependence of the total yield significantly diminishes. Recently, we observed CEP dependence in laser-induced acoustic waves and air fluorescence \cite{han2025hearing,liang2025waveform}. These novel CEP-dependent phenomena can be explained by the modulation of total ionization yield by the laser CEP. This provides an additional degree of freedom for controlling strong-field phenomena—beyond directional control—by linking the total ionization yield to the laser CEP.

Beyond the up-down asymmetry and total yield modulation, the holographic interference structures are also influenced by the laser CEP, as showed in supplement videos. The most common form of photoelectron holography is the so-called spider-leg structure \cite{huismans2011time,de2020all}, which arises from the interference between two electron pathways originating within the same quarter of the optical cycle. As illustrated in Fig. 1(g), one pathway corresponds to the direct electron (red arrow), while the other represents the forward-scattered electron (black arrow). The spider-leg holography structure has been utilized to probe molecular structures and investigate non-adiabatic sub-cycle electron dynamics \cite{zhou2016near,he2018direct,tan2018determination,cruz2023forward,werby2021dissecting,kang2020holographic,brennecke2020gouy}. In contrast, another type of holography arises from the interference between direct electrons and backward-scattered electrons, as illustrated in Fig. 1(h), forming a fishbone-like structure \cite{bian2011subcycle}. This interference structure commonly appears in theoretical calculations \cite{bian2012attosecond} but has been challenging to observe experimentally \cite{haertelt2016probing}. Our theoretical and experimental results both illustrate that the two types of interference structures can be separated in momentum space and therefore enhanced with a cosine-shaped single-cycle pulse.

\begin{figure}[htbp]
\centering
\includegraphics[width=\linewidth]{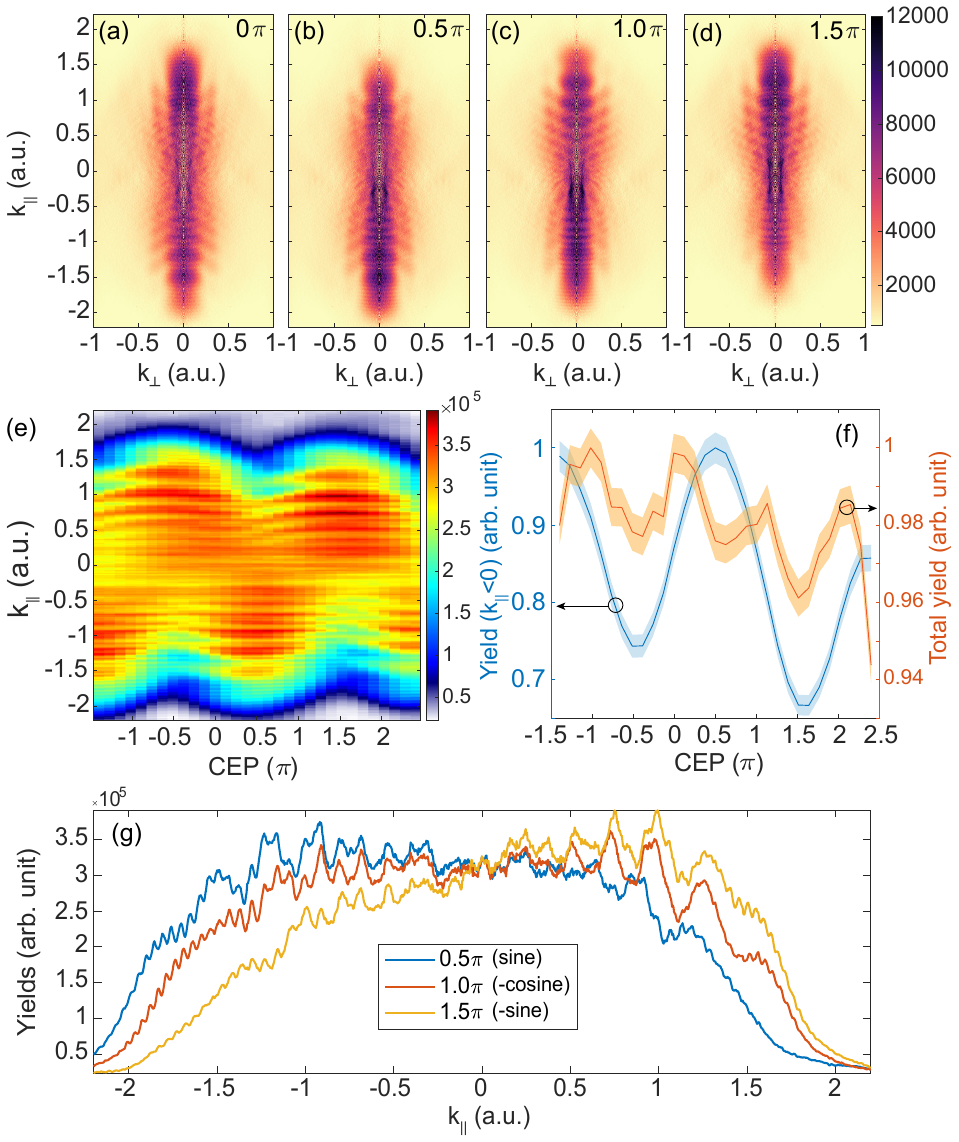}
\caption{(\textbf{a-d}) Measured photoelectron momentum distributions of Ar at CEP values of 0, 0.5, 1.0, and 1.5 $\pi$, respectively. (\textbf{e}) Py- and CEP-resolved photoelectron distribution of Ar, where Px is integrated from 0 to 0.25 a.u.. (\textbf{f}) CEP-resolved photoelectron yields from (e), obtained by integrating Py over the range from -2 a.u. to 0 a.u., as well as over the entire range. (\textbf{g}) Lineouts from (e) at CEP values of 0.5, 1.0, and 1.5 $\pi$. At the CEP value of 1.0 $\pi$ (red curve), the ATI contrast has the largest discrepancy between the up and down electrons, in agreement with the simulation in Fig. 1(e).}
\label{fig:figure2}
\end{figure}

Experimentally, we performed measurements using industrial-grade Yb-based laser pulses, post-compressed to 3.7 fs and characterized via TIPTOE and transient-grating FROG techniques \cite{han2025hearing,liang2025waveform}. A pair of 2.8-degree fused silica wedges was used for fine CEP tuning. The CEP stability was better than 200 mrad in 48 hours measured with a single-shot f-to-2f interferometer. A VMI apparatus recorded two-dimensional photoelectron momentum projections. The exposure time  of the VMI camera was set to 0.1 s per frame. At each CEP position, 1000 frames were averaged. The Abel inversion algorithm of BASEX \cite{dribinski2002reconstruction} was applied to extract the central momentum slice. 

Figures 2(a-d) display the Abel-inverted PMDs of Ar at CEP values of 0, 0.25, 0.5, and 0.75 $\pi$. Supplementary videos provide CEP-dependent PMDs for Ar and N$_2$ molecules. From Figures 2(a–d) and the videos, we identify at least two key features. First, the intensity-weighted momentum center shifts up and down, consistent with $k_f = -A(t_0)$, where $k_f$ is the photoelectron momentum and $A(t_0)$ is the laser vector potential at the peak instant $t_0$. Second, the contrast of inter-cycle and intra-cycle interference fringes differs between upward and downward electrons, both exhibiting strong CEP dependence. 

Figure 2(e) shows the PMD projected onto the laser polarization axis as a function of CEP. Figure 2(g) extracts several lineouts from Fig. 2(e) at CEP values of 0.5, 1.0, and 1.5 $\pi$. At CEP = 1.0 $\pi$, corresponding to a negative cosine-shaped pulse, the intensity-weighted center of the distribution lies close to zero; however, the inter-cycle fringe contrast is markedly stronger in the downward direction ($k_{||}<0$) than in the upward direction, in excellent agreement with our TDSE results in Fig. 1(e). At CEP = 0.5 $\pi$ (sine-shaped pulse), the intensity-weighted center of the PMD shifts maximally downward, whereas at CEP = 1.5 $\pi$ (negative sine-shaped pulse) it shifts maximally upward. This behavior arises because the vector potential of a sine-shaped pulse is cosine-shaped, leading to the strongest up–down asymmetry, consistent with the streaking relation $k_f = -A(t_0)$.

Figure 2(f) presents the photoelectron yields as a function of CEP in both the half-momentum space (i.e., $k_{||}<0$) and the entire momentum space. The half-momentum-space yield exhibits a CEP dependence with a periodicity of 2$\pi$, where maxima (minima) correspond to sine- (negative sine-) shaped pulses. In contrast, the total yield follows a CEP dependence with $\pi$ periodicity, reaching its maximum at cosine-shaped and negative cosine-shaped pulses. N$_2$ shows very similar results as Ar. These experimental results align well with the TDSE and ADK model predictions in Fig. 1(f). Note that the slight non-periodicity along the CEP axis is due to the pulse duration stretching effect by varying the thickness of fused silica inserted into the beam path. 

Having established accurate experimental laser pulse parameters, we now proceed to answer our main question: what target structure information can we retrieve from the holographic patterns in the measured PMD? It has been demonstrated theoretically \cite{morishita2008accurate,Chen:pra2009,Xu:pra2010} and experimentally \cite{blaga2012imaging,wolter2016ultrafast} in laser-induced electron diffraction (LIED) that the high-energy region of the PMD, originated from rescattered electrons, can be used to extract laser-free electron-target ion elastic scattering differential cross section (DCS) at large scattering angles. The DCS can then be used to retrieve time-resolved molecular structure information such as bond lengths and bond angles of the target at the moment of rescattering, with a sub-femtosecond temporal resolution \cite{morishita2008accurate,blaga2012imaging,wolter2016ultrafast}. The main drawback of the LIED technique is that the signals at the high-energy region are very weak -- significantly weaker than in the holographic region. It has been therefore anticipated since the pioneering work by Huimans {\it et al}  \cite{huismans2011time} that this strong-field holographic electron imaging technique can provide an alternative tool for dynamic imaging of molecular structure. In fact, significant progress has been reported \cite{haertelt2016probing,zhou2016near,khurelbaatar2024strong,he2018direct,tan2018determination,kang2020holographic,Huismans:prl2012, Hickstein:prl2012,bian2012attosecond, Meckel:NaturePhys2014,Liu:prl2016, Walt:NatureCom2017,Tan:pra2019,Li:prl2019,Xie:prl2021,Min:OptExp2024,de2020all,cruz2023forward,werby2021dissecting,brennecke2020gouy}. An important step forward was the theoretical method proposed by Zhou {\it et al} \cite{zhou2016near}, which showed that the phase of the scattering amplitude can, in principle, be extracted from holographic photoelectron spectra. Here we propose a modified approach and demonstrate that an accurate {\it laser-free} scattering amplitude phase can be retrieved from experimentally measured PMD. 


In Fig.~3(a), we compare simulated PMD with the experimental measurement for Ar at CEP=0. Our TDSE simulation nicely reproduces all the holographic features of the experiment, including the spider-leg and the fishbone structures, mostly visible in the upper half and the lower half of the PMD, respectively. Note that to get such a good agreement with the experiment, averaging over the focal volume intensity variation was carried out for the peak intensity of $2.8 \times 10^{14}$ W/cm$^2$ (see Supplemental Material for more details). In the following, we will analyze the upper half of the PMD, in which the spider-leg structure resulting from the interference between near-forward rescattered electron and direct electron emitted in the same sub-cycle at the peak of the cosine-shaped pulse dominates, see Fig.~1(g). To make the comparison more quantitative, we show the zoom-in part of the measured PMD together with the local minimum positions of the yields from both experiment and simulation, see Fig.~3(b).  To extract the near-forward scattering phase from these interference patterns, we first analyze the yields at a slice at a constant $k_{\parallel}=0.9$ a.u.. After subtracting a smooth monotonically decreasing background, the decaying experimental signal vs $k_{\perp}$ is obtained and fitted to a smooth oscillating function, see Fig.~3(c). Dividing this signal by a smooth envelope, we then get an oscillating function bounded in the range between -1 and +1, shown in Fig.~3(d) together with the TDSE result. 

\begin{figure}[htbp]
\centering
\includegraphics[width=\columnwidth]{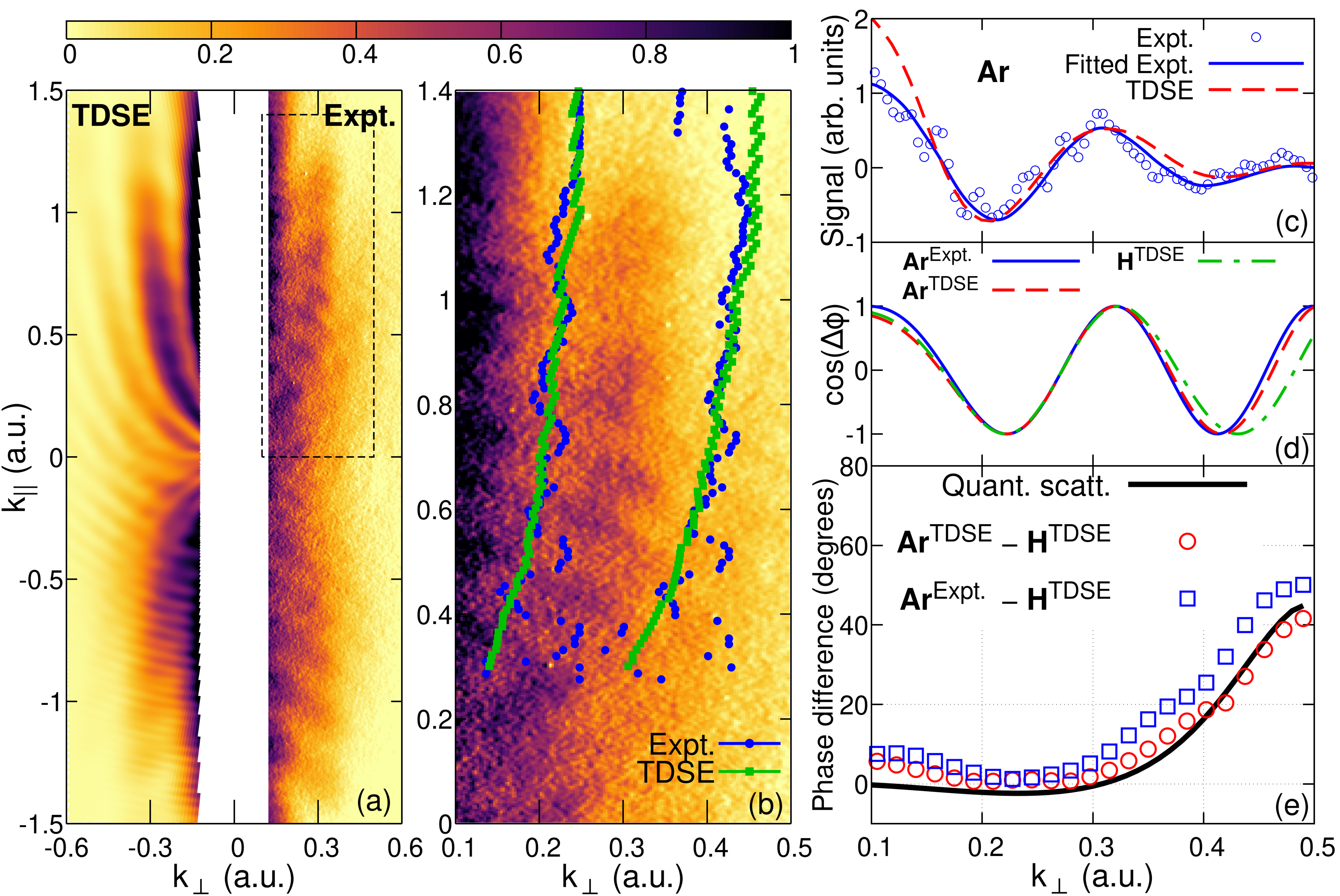}
\caption{(\textbf{a}) Side-by-side comparison of PMDs from TDSE and experiment for Ar at CEP=0. The central parts are removed due to the artifacts from VMI Abel inversion. (\textbf{b}) Zoomed-in experimental results with the extracted minimum positions from TDSE and experiments. (\textbf{c}) Signals vs $k_{\perp}$ after a smooth background is subtracted at $k_{\parallel}=0.9$. (\textbf{d}) Same as (\textbf{c}), but after further division by a smooth envelope. (\textbf{e}) Retrieved phase of the {\it laser-free} scattering amplitude from Ar relative to that of hydrogen atom. Result from standard quantum scattering theory is also shown.}
\label{fig:figure3}
\end{figure}

\begin{figure}[htbp]
\centering
\includegraphics[width=0.8\linewidth]{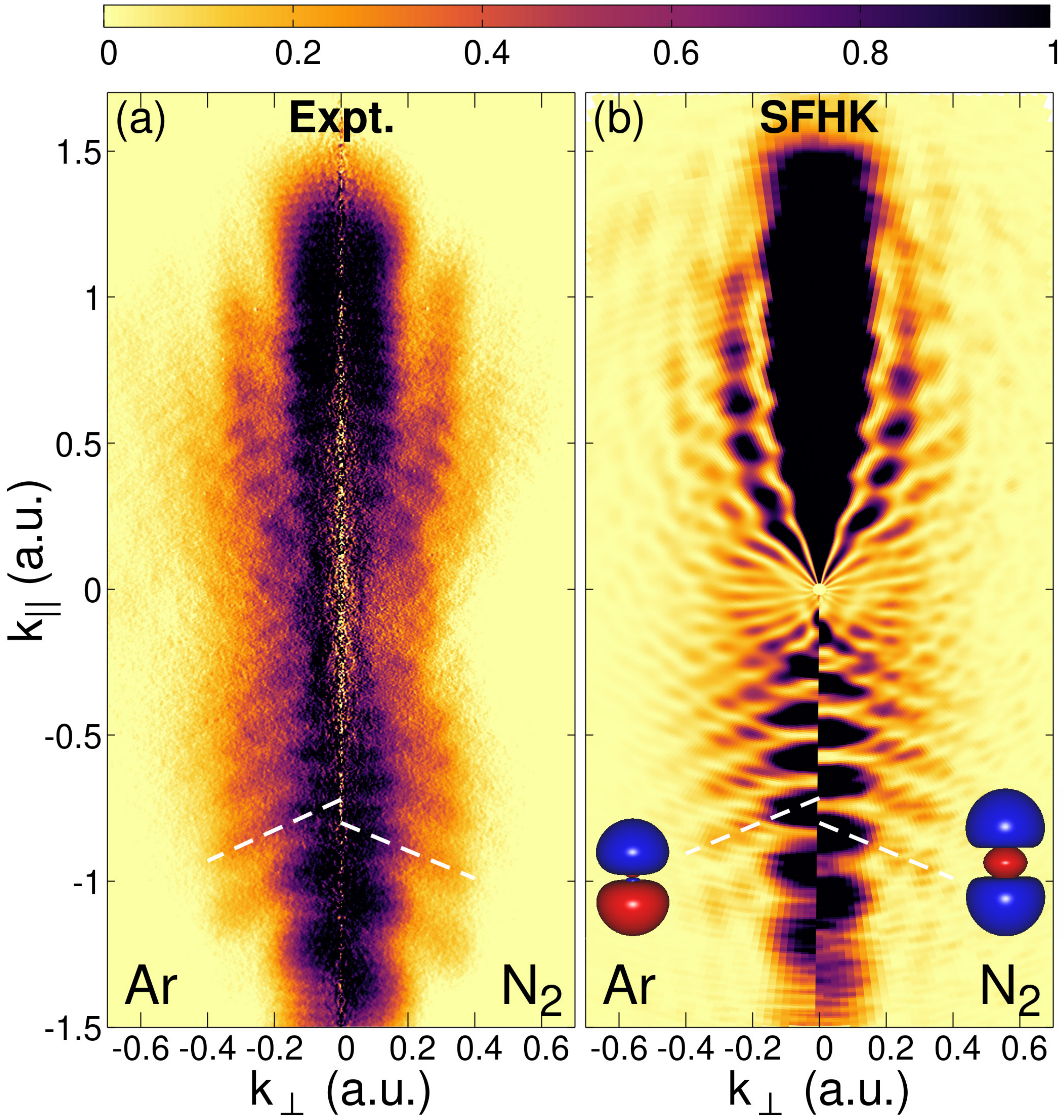}
\caption{(\textbf{a}) Side-by-side comparison of measured PMDs of Ar (left) and N$_2$ (right) at the same laser intensity with CEP=0. (\textbf{b}) The corresponding comparison from our SFHK simulation for the same laser parameters as of the experiment.}
\label{fig:figure4}
\end{figure}

The above procedure has been proposed by Zhou {\it et al} \cite{zhou2016near}, together with their mapping method, to extract the scattering amplitude phase for short-range potentials (without a Coulomb tail). Furthermore, their method was applied to the case of a single laser intensity, in which, for any given point $(k_{\perp},k_{\parallel})$, the ionization and rescattering times can be calculated uniquely for each subcycle. To proceed further with real experimental data, we therefore propose the following modifications. First, we use a reference atomic target (taken to be hydrogen) so that only the phase difference between our target and hydrogen needs to be retrieved. It is expected that the corrections due to the influence of the Coulomb tail will largely be canceled out in the retrieval, as the same Coulomb tail affects both targets. Second, we define the ``effective" laser intensity to be the median intensity in the differential intensity contribution to the theoretical PMD. This effective intensity is used for our mapping (see Supplemental Material for more details). 

Due to the lack of experimental data for the reference hydrogen atom, in our retrieval procedure we will use the simulated TDSE result of hydrogen with the same laser parameters as for our target. The phase retrieved from experimental data for Ar relative to that of hydrogen is shown in Fig.~3(e). The same procedure was also applied to the simulated data. For comparison, we also show in Fig.~3(e)  the phase difference calculated with standard quantum scattering theory. This successful retrieval of the phase of scattering amplitude can be seen for other cuts at different $k_{\parallel}$. The phase difference is quite small at $k_{\perp} \lesssim 0.3$, reflecting the fact that small-angle scattering is dominated by the long-range Coulomb potential. It increases with $k_{\perp} \gtrsim 0.3$, indicating more influence of the short-range potential, as the electron penetrates closer to the core. We remark that the use of an ultrashort pulse with a well-controlled CEP helps better separation of the spider-leg structures from the fishbone-like structures that simplifies our procedure and improves the quality of our retrieved phase. 

Finally, we remark that the parity of atomic and molecular orbitals of the targets can be detected by analyzing the holographic patterns. This has been demonstrated in fine structures in the direction perpendicular to the light polarization by Kang {\it et al} \cite{kang2020holographic}. With our ultrashort, CEP-controlled pulses, the influence of parity can be seen even more clearly. Our experimental PMDs from Ar and N$_2$ at the same light intensity for CEP=0 are compared side-by-side in Fig.~4(a). The signals from N$_2$ show clear out-of-phase patterns as compared to those from Ar. These striking out-of-phase features are reproduced in our simulations based on the SFHK method \cite{tran2024quantum}, see Fig.~4(b). Note that all these patterns survive from focal volume intensity averaging, but they become less pronounced. The out-of-phase character of these two targets can be traced back to the different parities of the atomic or molecular orbital of the active electron, which is odd for Ar(3p) and even for N$_2(\sigma_g)$. For example, we have checked that if an $s$ orbital is used in our simulation for Ar(3p), an in-phase pattern with N$_2$ would be seen. Note that in the simulation for N$_2$, the laser polarization is assumed to be parallel to the molecular axis. We found that the out-of-phase patterns do not change in the perpendicular case. We remark that recently Khurelbaatar {\it et al} \cite{khurelbaatar2024strong} used the fishbone-like structures to retrieve N$_2$ bond length with a shorter wavelength (723 nm) Ti:Sa laser.

In summary, we have demonstrated strong-field interferometry with CEP-stabilized, near-single-cycle Yb lasers, achieving high-resolution measurements of photoelectron momentum distributions from atoms and molecules. The use of industrial-grade Yb sources, combined with post-compression and excellent CEP stability, enabled us to overcome long-standing limitations of few-cycle Ti:Sa lasers and to observe CEP-dependent holographic structures with unprecedented clarity. We showed that cosine-shaped pulses separate and enhance both spider-leg and fishbone holography, allowing direct access to the electron scattering phase and to orbital parity signatures in atomic and molecular targets. The retrieved scattering phase agrees quantitatively with \textit{ab initio} theory and validates the recently developed SFHK method, establishing holographic interferometry as a practical route to precision measurements of electron–molecule scattering in the strong-field regime. Our results open the way toward dynamic molecular imaging with low-energy electrons, using compact and robust laser technology.

We thank C. Aikens, J. Millette and S. Chainey for their technical support, and E. Mullins for sharing his expertise on VMI operation. The Kansas group was supported by the Chemical Sciences, Geosciences and Biosciences Division, Office of Basic Energy Sciences, Office of Science, US Department of Energy, Grant No. DE-FG02-86ER13491. Theoretical simulations by P.H.T. and A.T.L. were supported by DOE BES, Chemical Sciences, Geosciences, and Biosciences Division Grant No. DE-SC0023192. M.-C.C. thanks the National Science and Technology Council, Taiwan, for funding grant no. 113-2112-M-007-042-MY3.

\bibliography{pop_references}

\end{document}